\documentclass{aa}  

\usepackage{graphicx}
\usepackage{txfonts}
\usepackage{tablefootnote}
\usepackage{hyperref}

\usepackage{xcolor}
\definecolor{dark-red}{rgb}{0.9,0.0,0.0}
\definecolor{dark-blue}{rgb}{0.15,0.15,0.9}
\definecolor{dark-green}{rgb}{0.15,0.8,0.15}
\definecolor{medium-blue}{rgb}{0,0,0.9}
\hypersetup{
    colorlinks, linkcolor=red,
    citecolor={dark-blue} , urlcolor={medium-blue}
}
\usepackage{changepage} 

\newcommand{\mjup}{$M_{\rm J}$\,}
\newcommand{\msun}{$M_\odot$\,}
\newcommand{\lsun}{$L_\odot$\,}

\newcommand{\rsun}{$R_\sun$\,}
\newcommand{\mstar}{$M_\star$\,}
\newcommand{\rstar}{$R_\star$\,}
\newcommand{\logg}{log\,$g$\,}
\newcommand{\feh}{{\rm [Fe/H]}\,}
\newcommand{\vsini}{\textit{v}\,sin\,$i$}
\newcommand{\ms}{m\,s$^{-1}$\,}
\newcommand{\cmsq}{cm\,s$^{-2}$\,}
\newcommand{\teff}{$T_{{\rm eff}}$\,}
\newcommand{\mplan}{m$_b$\,sin$i$\,}
\newcommand{\gaia}{{\small\texttt{Gaia}}\,}
\newcommand{\hipparcos}{{\small\texttt{Hipparcos}}\,}
\newcommand{\msyr}{m\,s$^{-1}$\,yr$^{-1}$}

\newcommand{\mm}{\,$\pm$\,}

\newcommand{\plansix}{HIP\,114933\,$b$}

\begin{document}

   \title{Four Jovian planets around low-luminosity giant stars observed by the EXPRESS and PPPS\thanks{ Based on observations collected at La Silla - Paranal Observatory under programs ID's 085.C-0557, 087.C.0476, 089.C-0524, 090.C-0345 and through the Chilean Telescope Time under programs ID's CN-12A-073, CN-12B-047, CN-13A-111, CN-2013B-51, CN-2014A-52, CN-15A-48, CN-15B-25 and CN-16A-13.}}


   \author{ M. I. Jones \inst{1}
    \and R. Wittenmyer \inst{2}
    \and C. Aguilera-G\'omez\inst{3}
    \and M. G. Soto \inst{4}
    \and P. Torres \inst{5}
    \and T. Trifonov \inst{6}
    \and J. S. Jenkins \inst{7,8}
    \and A. Zapata \inst{5}
    \and P. Sarkis \inst{6}
    \and O. Zakhozhay \inst{6,9}  
    \and R. Brahm \inst{10,11}
    \and F. Santana \inst{7}
    \and J. I. Vines \inst{7}
    \and M. R. D\'iaz \inst{7}
    \and M. Vu\v{c}kovi\'{c} \inst{12}
   }

   \institute{European Southern Observatory, Alonso de C\'ordova 3107, Vitacura, Casilla 19001, Santiago, Chile \\\email{mjones@eso.org}
   \and  University of Southern Queensland, Centre for Astrophysics, Toowoomba, QLD 4350, Australia
   \and Departamento de Ciencias Físicas Universidad Andrés Bello. Fern\'andez Concha 700, 759-1538, Las Condes, Santiago, Chile.
   \and School of Physics and Astronomy, Queen Mary University, 327 Mile End Road, London E1 4NS, UK
   \and Instituto de Astrof\'isica, Facultad de F\'isica, Pontificia Universidad Cat\'olica de Chile, Av. Vicu\~na Mackenna 4860, 7820436 Macul, Santiago, Chile
   \and Max-Planck-Institut f\"{u}r Astronomie, K\"{o}nigstuhl 17, D-69117 Heidelberg, Germany
   \and Departamento de Astronom\'ia, Universidad de Chile, Camino El Observatorio 1515, Las Condes, Santiago, Chile
   \and Centro de Astrof\'isica y Tecnolog\'ias Afines (CATA), Casilla 36-D, Santiago, Chile
    \and Main Astronomical Observatory, National Academy of Sciences of the Ukraine, 03143 Kyiv, Ukraine
    \and Facultad de Ingenier\'ia y Ciencias, Universidad Adolfo Ib\'a\~nez, Av.\ Diagonal las Torres 2640, Pe\~nalol\'en, Santiago, Chile
    \and Millennium Institute for Astrophysics, Chile
   \and Instituto de F\'isica y Astronom\'ia, Universidad de Valpara\'iso, Casilla 5030, Valpara\'iso, Chile 
}

   \date{}

 
  \abstract{We report the discovery of planetary companions orbiting four low-luminosity giant stars with \mstar between 1.04 and 1.39 \msun. All four host stars have been independently observed by the EXoPlanets aRound Evolved StarS (EXPRESS) program and the Pan-Pacific Planet Search (PPPS). The companion signals were revealed by multi-epoch precision radial velocities obtained during nearly a decade. The planetary companions exhibit orbital periods between $\sim$ 1.2 and 7.1 years, minimum masses of m$_{\rm p}$sin\,$i$ $\sim$ 1.8\,-\,3.7 \mjup and eccentricities between 0.08 and 0.42. Including these four new systems, we have detected planetary companions to 11 out of the 37 giant stars that are common targets between the EXPRESS and PPPS. After excluding four compact binaries from the common sample, we obtained a fraction of giant planets (m$_{\rm p} \gtrsim$ 1\,-\,2 \mjup) orbiting within 5 AU from their parent star of $f = 33.3^{+9.0}_{-7.1} \%$. This fraction is significantly higher than that previously reported in the literature by different radial velocity surveys. Similarly, planet formation models under predict the fraction of gas giant around stars more massive than the Sun. 
 }


   \maketitle
%

\section{Introduction}

After the discovery of the first extrasolar planet around a solar-type star \citep{mayor1995}, thousands of new planetary companions have been uncovered. During the 90's and 2000's, hundreds of them were discovered  using the radial velocity (RV) method. More recently, with the advent of dedicated space missions, like Kepler \citep{kepler} and the Transiting Exoplanet Survey Satellite \citep[TESS;][]{tess}, the number of newly detected planets has increased by almost an order of magnitude thanks to the exquisite photometric precision achieved by these missions capable to measure transit depths as small as $\sim$200 ppm (e.g. \citealt{jenkins2015}). Using this information, we have been able to compute occurrence rates of short-period planets around low-mass M-dwarfs (\citealp{mulders2015,hardegree2019}) and Sun-like stars (\citealp{petigura2013,barbato2018}). In addition, by combining the transit information with dedicated ground-based RV follow-up, it is possible to fully characterize the physical properties of transiting planetary companions (planet radius, mass and density; e.g. \citealp{jones2019}), allowing us to study their internal structure and composition (e.g. \citealp{thorngren2016}).
Moreover, long-running RV surveys searching for planetary companions to solar-type stars have provided a detailed understanding of the planet population at short and large orbital separations (e.g. \citealp{marcy2000,wittenmyer2020a}). \newline \indent
On the other hand, intermediate-mass (IM) stars (\mstar $\gtrsim$ 1.5\,\msun) have been mainly excluded from long-term RV surveys because of the inherent difficulties to measure precision RVs from their optical spectra (e.g. \citealp{lagrange2009}), which restricts the amplitudes of the RV signals to the massive companions regime. It is possible however to search for planets orbiting such massive stars via precision RVs by studying their post-main-sequence descendants (e.g. \citealt{johnson2007a}). For more than 15 years different RV surveys have targeted evolved stars, aimed at computing occurrence rates and thus to establish the role of the stellar mass in planetary systems around IM star (\citealp{frink2001,setiawan2003, sato2005,hatzes2005,niedzielski2007,han2010,jones2011,wit2011}). These surveys have detected $\gtrsim$ 150 planets and have found interesting correlations between the frequency of giant planets and the stellar mass and metallicity (\citealp{reffert2015,jones2016,wit2017a}) and a lack of short-period gas giants (\citealp{sato2008,dollinger2009}). \newline \indent
In this paper we present radial velocity measurements of four giant stars that have been targeted by the EXoPlanets aRound Evolved StarS (EXPRESS; \citealt{jones2011}) and the Pan-Pacific Planet Search (PPPS; \citealt{wit2011}). The obtained RVs of these stars revealed periodic variations, which are most likely attributed to the presence of companions in the planetary mass regime.
The orbital period of the planetary companions cover a range from 1.2 to 7.1 years, and projected companion masses between $\sim$ 1.8\,-\,3.7 \mjup. 
The paper is organized as follows: the observations, data reduction and RV measurements are presented in section \ref{sec2}. In section \ref{sec3} we present the host star properties. The RV analysis and orbital fitting is presented in section \ref{sec4}. In section \ref{sec:detectability} we computed planet detectability curves for all four host stars. In section \ref{sec:stellar_activity} we study the stellar activity indicators to discard false positive scenarios and in section \ref{sec:astrometry} we use
the available Hipparcos data to search for possible astrometric orbits induced by the companions. Finally the summary and discussion is presented in section \ref{sec:conclusions}.

\section{Observations \label{sec2}}

High-resolution spectra have been taken for the four giant stars presented here. These observations have been obtained as part of the EXPRESS and PPPS surveys. 
These two programs share a total of 37 targets, and several planets have been published using their combined data (\citealt{jones2016}; \citealt{wit2016a}; \citealt{wit2017a}).
A description of the observational strategy and data analysis of these two programs is given in the next sections.

\subsection{EXPRESS data}

In 2009 we began the observing campaign of the EXPRESS targets, whose sample consists of a total of 166 bright stars (V $<=$ 8 mag), which
are observable from Chile (dec $<$ 20 deg). The target selection criteria are explained in \citet{jones2011}.
In total, we have obtained $\gtrsim$ 15 spectra per star, with a typical time baseline of $\gtrsim$ 5 years. Given our observational cadence and number of observations, we can 
efficiently detect planetary companions inducing RV amplitude of $\gtrsim$ 25 \ms, with orbital period up to several years.
The first EXPRESS observations were performed in 2009 with the fiber echelle (FECH) spectrograph at the 1.5m telescope, placed in the Cerro Tololo Inter-American Observatory (CTIO). Unfortunately, even though some of these observations were used to detect a few large amplitude signals (e.g. \citealt{jones2014}), most of these observations were discarded, since the lack of thermal and mechanical stability and relatively low resolution of the instrument, precluding us from computing RVs with a long-term precision better than $\sim$\,20-30 \ms. However, in 2011 FECH was replaced by a much higher resolution and more stable instrument called CHIRON \citep{TOK13}, which we used until 2016, and then again in 2019. These observations were also complemented since 2010 with the Fiber-fed Extended Range Optical Spectrograph (FEROS; \citealt{kaufer1999}). 
We note that both CHIRON and FEROS observations deliver good quality data, leading to a long-term precision of $\sim$ 5 \ms, which is well suited for this program. For the data reduction and Doppler analysis, we have developed several automatic tools, that have been described extensively in different papers (\citealp{jones2017}; \citealp{jones2018}), and are briefly described below.
The FEROS data are reduced with the CERES code \citep{brahm2017}, which does an optimal extraction of the echelle orders individually, after
applying the detector bias corrections. The orders are traced and extracted using a combined flat-field spectrum, taken before the start of the science observations. Finally, the wavelength solution is applied to both the science and calibration fiber. Similarly, the CHIRON data are extracted and calibrated using the Yale pipeline \citep{TOK13}.
For the FEROS data, we measured the radial velocities using the cross-correlation technique and the simultaneous calibration method (\citealp{baranne1996}).
Briefly, we compute the cross-correlation-function (CCF) between a high signal-to-noise template of the same star. The template is created by stacking all of the FEROS observations for each star, after correcting by their relative Doppler shift. To correct for the nightly spectral drift, we subtract the velocity that is measured 
from the simultaneous calibration lamp, which is recorded in the coupled-charge device (CCD) thanks to the use of the simultaneous calibration fiber. This method is repeated independently in four chunks per each of the 25 orders, covering the wavelength range between $\sim$ 3900\,\AA\,-\,6800\,\AA\, (more details can be found \citealt{jones2017}). The resulting velocity is computed from the median of 100 individual chunk velocities. The corresponding RV uncertainties are computed from the error in the mean of all 100 chunk velocities. The FEROS RVs and uncertainties are listed in the online Tables \ref{hip56640_rv}-\ref{hip114933_rv}. In addition, the pipeline also computes the bisector velocity span (BVS; \citealt{queloz2001}), and activity indicators based on the Mount Wilson system \citep{duncan1991}, following the method described in \citet{jenkins2008} and \citet{jones2017}.  
We note that we have inflated the internal Smw errorbars by a factor 0.0037, which is added in quadrature. This value corresponds to the standard deviation of the S-index obtained from 85 FEROS spectra of the quiet RV standard star $\tau$ Ceti.
\newline \indent
CHIRON is equipped with an I$_2$ cell, which superimposes a dense absorption spectrum on top of the stellar 
light that is recorded in the CCD. To obtain the velocities from the data, we use the I$_2$ cell method, following the prescription
presented in \cite{butler1996}, but using only one Gaussian to model the instrumental profile. We apply this method to $\sim$ 3\,\AA\, chunks
in 22 different orders, covering the wavelength range between $\sim$ 5000\,-\,6200\,\AA. The final RVs are 
obtained from the weighted mean individual chunk velocities. The individual statistical weights are computed from the long-term scatter of each individual chunk velocity. More details can be found in \cite{jones2017}.
The CHIRON RVs and uncertainties are listed in the online Tables \ref{hip56640_rv}-\ref{hip114933_rv}.

\subsection{PPPS data}

The Pan-Pacific Planet Search (PPPS) originated in 2009 as a Southern hemisphere 
extension of the established Lick \& Keck Observatory survey for planets 
orbiting Northern ``retired A stars'' \citep[e.g.][]{johnson2006b, johnson2007a, johnson2010}.  Data from the PPPS have contributed to the discovery of 16 planets, including 8 planets in collaboration with EXPRESS.  Notable discoveries include the 3:5 resonant pair of giant planets orbiting HD\,33844 \citep{wit2016a} and HD\,76920\,$b$, the most eccentric planet ever found to orbit an evolved star \citep{wit2017b,bergmann2020}.  We obtained observations between 2009 and 2015 using the UCLES high-resolution spectrograph \citep{diego90} at the 3.9-metre Anglo-Australian Telescope (AAT).  UCLES achieves a resolution of 45,000 with a 1-arcsecond slit.  An iodine absorption cell provides wavelength calibration from 5000 to 
6200\,\AA.  Precise radial velocities are determined using the iodine-cell technique as noted above and as detailed in \citep{butler1996}.  The photon-weighted mid-time of each exposure is determined by an exposure meter. All velocities are measured relative to the zero-point defined by the iodine-free template observation.  
The UCLES velocities are given in Tables \ref{hip56640_rv}\,- \,\ref{hip114933_rv}.  

\section{Stellar properties \label{sec3}}

\begin{figure}
\includegraphics[angle=0,scale=0.39]{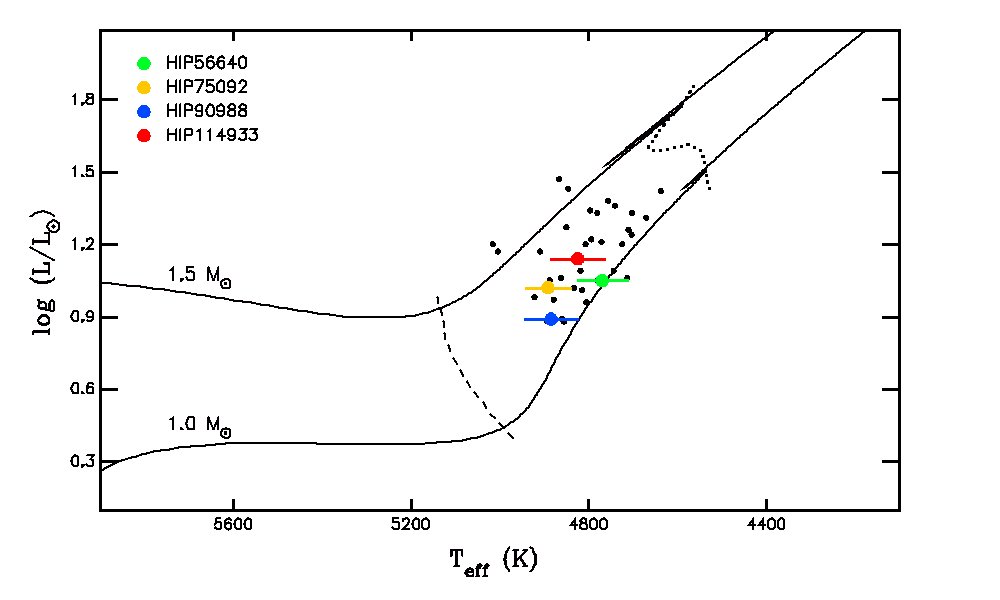}
\caption{HR diagram showing the position of the 4 host stars (fill circles). The small black dots represent the 37 common targets between the EXPRESS and PPPS. The solid lines correspond to 1.0 \msun and 1.5 \msun \texttt{PARSEC} evolutionary tracks at solar metallicity \citep{bressan2012}. The dashed line represents the base of the RGB, while the dotted line corresponds to the lower luminosity bump region.\label{hrdiagram} }
\end{figure}

The stellar properties of the studied targets are summarized in Table \ref{stellar_par}. The $BVJHK$ magnitudes were retrieved from the Tycho-2 catalogue \citep{hog2000} and the 2-MASS All-Sky catalogue \citep{cutri2003}, while the spectral types from the Michigan Catalogue of two-dimensional spectral types for the HD stars (\citealt{houk1978}; \citealt{houk1982}; \citealt{houk1988}).
The interstellar visual absorption coefficients (A$_V$) were obtained using the dust extinction maps presented in \citet{bovy2016}.
The distance to the host stars was computed from the Gaia DR2 \citep{gaiadr2} parallaxes, after correcting by the systematic offset 
found by \citet{stassun2018}. In addition, we recomputed the atmospheric parameters, using an updated version of the SPECIES code (\citealt{species}; Soto et al. in preparation). 
For this purpose, we used a high signal-to-noise-ratio (SNR) template, which was built by stacking all of the individual observed FEROS spectra for each star. These new high SNR templates lead to internal uncertainties smaller than those presented in \citet{jones2011}. The resulting \teff, \logg and [Fe/H] values are also listed in Table \ref{stellar_par}. Figure \ref{hrdiagram} shows the resulting position of the four stars in the HR diagram (filled circles). For comparison the position of all 37 targets in common to the EXPRESS and PPPS are overplotted (small black dots). As can be seen all four host stars are first ascending the red giant branch (RGB) phase, and are located in between the RGB base (dashed line) and the luminosity bump. Since no horizontal branch (HB) evolutionary track crosses this region, we can safely identify these stars as first ascending RGB, and thus we can derive their ages in a more accurate way when compare to clump giants. \newline
Finally, the stellar mass, radius, luminosity, and age of the host stars were estimated using the latest version of the \texttt{isochrones}\footnote{\url{https://github.com/timothydmorton/isochrones}} package \citep{Morton2015}. For this, we compared the atmospheric parameters derived with SPECIES to a grid of MESA Isochrones and Stellar Tracks \citep{Dotter2016} using the MultiNest tool \citep{Feroz2009}. Other input quantities are the $BVJHK$ magnitudes and the gaia DR2 parallax. The resulting stellar physical parameters are listed in Table \ref{stellar_par}, and are those adopted for the rest of the paper. We note that the stellar masses presented here are systematically smaller than those presented in \citet{jones2011} by -0.35 \msun. We also find that the newly computed distances and stellar luminosities are more accurate and in some cases might largely depart from those presented in \cite{jones2011}.
The main reasons for these differences is due to a combination of lower effective temperature found here, differences in the \gaia parallaxes compared to the \hipparcos ones \citep{hipparcos2007}, and different stellar evolutionary tracks used in the two studies. 

\begin{table*}
\centering
\caption{Stellar properties \label{stellar_par}}
\begin{tabular}{lllll}
\hline\hline
\vspace{-0.3cm} \\
                    & HIP\,56640              & HIP\,75092               & HIP\,90988                & HIP\,114933     \\
\hline \vspace{-0.3cm} \\
Spectral Type      & K1III                    & K0III                    &  K1III                    & K0III           \\ 
$B$ (mag)          & 9.02$\pm$0.02            & 8.15$\pm$0.02            & 8.81$\pm$0.02             & 8.29$\pm$0.02    \\
$V$ (mag)          & 7.93$\pm$0.01            & 7.11$\pm$0.01            & 7.76$\pm$0.01             & 7.25$\pm$0.01    \\
$J$ (mag)          & 6.133$\pm$0.017          & 5.358$\pm$0.019          & 5.995$\pm$0.018           & 5.564$\pm$0.018  \\ 
$H$ (mag)          & 5.639$\pm$0.024          & 4.935$\pm$0.023          & 5.531$\pm$0.031           & 5.049$\pm$0.020  \\
$K$ (mag)          & 5.493$\pm$0.017          & 4.738$\pm$0.027          & 5.383$\pm$0.020           & 4.907$\pm$0.016  \\
$\Pi$ (mas)        & 8.146$\pm$0.039          & 12.216$\pm$0.061         & 10.606$\pm$0.043          & 9.798$\pm$0.051  \\
\teff (K)          & 4769$\pm$55              & 4891$\pm$50              & 4884$\pm$58               & 4824$\pm$60      \\
$L_\star$ (\lsun)       & 11.22$^{+0.26}_{-0.26}$   & 10.47$^{+0.34}_{-0.26}$  & 7.76$^{+0.18}_{-0.17}$    & 13.80$^{+0.26}_{-0.0.22}$\\
\logg (\cmsq)      & 2.91$\pm$0.12            & 3.09$\pm$0.10            & 3.32$\pm$0.11             &  2.99$\pm$0.12   \\
\feh (dex)         & -0.03$\pm$0.05           & -0.01$\pm$0.05           & +0.17$\pm$0.06            & +0.06$\pm$0.06   \\
\vsini\, (k\ms)    & 3.1$\pm$0.7              & 3.1$\pm$0.5              & 3.5$\pm$0.6               & 3.3$\pm$0.5      \\
$M_\star$(\msun)   &  1.04$^{+0.07}_{-0.06}$  & 1.28$^{+0.11}_{-0.10}$   & 1.30$^{+0.08}_{-0.08}$    & 1.39$^{+0.09}_{-0.09}$   \\
\rstar(\rsun)      &  4.93$^{+0.05}_{-0.05}$  & 4.53$^{+0.05}_{-0.05}$   & 3.94$^{+0.04}_{-0.04}$    & 5.27$^{+0.05}_{-0.05}$   \\
A(Li) (dex)        &  <0.03                   &    <0.30                 &    <0.35                  & <0.29            \\
C$^{12}$/\,C$^{13}$& >40                      & 25.4$\pm$5.0             & 36.2$\pm$6.8              &   27.2$\pm$  6.0  \\
\vspace{-0.3cm} \\\hline\hline
\end{tabular}
\end{table*}

\subsection{Asteroseismology}

\begin{figure}
\includegraphics[angle=0,scale=0.39]{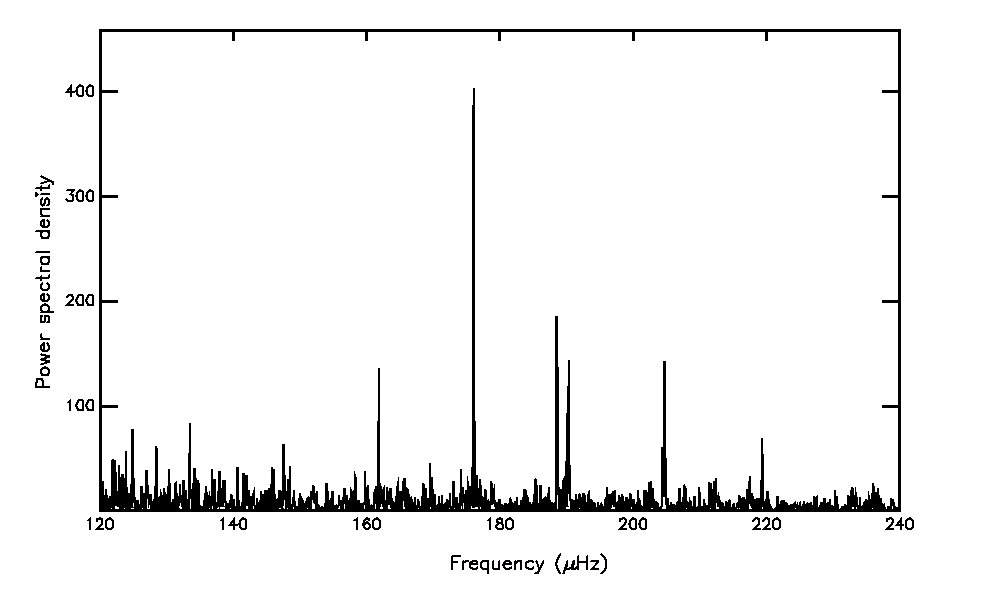}
\caption{Background-corrected PSD around the asteroseismic power excess region for HIP\,75092. \label{PSD}}
\end{figure}

In addition to the spectroscopic analysis, we performed an asteroseismic analysis of the host stars to derive their physical properties using well established scaling relations \citep[e.g. ][]{kjeldsen1995,stello2017}. For this purpose we first analyzed the TESS light curves of all four stars. Unfortunately, all of them were observed in the long-cadence mode ($\Delta t$ $\sim$ 30 minutes) making very difficult to detect the asteroseismic signals. However, HIP\,75092 was also observed by the $K2$ mission \citep{kepler2} in the short-cadence mode ($\Delta t \sim$ 1 min) in sector 15. 
The $K2$ light curve\footnote{The corrected light curve was retrieved from the MAST portal: \url{https://mast.stsci.edu/portal/Mashup/Clients/Mast/Portal.html}.} contains a total of 129239 individual measurements collected during 88 days. Before analyzing this dataset, we first corrected the light curve using a linear fit and we removed outliers using an iterative sigma clipping rejection algorithm.
Using the corrected light curve, we computed a generalized Lomb-Scargle periodogram \citep[GLS;][]{Zechmeister2009} with the \texttt{astropy.stats} Python module to obtain the power spectral density (PSD), from which we can measure the frequency of maximum power ($\nu_{\rm max}$) and the large frequency separation ($\Delta\nu$). After visual inspection of the PSD we could clearly identify the region of power excess with a high-significance peak around $\sim$ 180 $\mu$Hz (see Figure \ref{PSD}). 
We then corrected the background of the PSD in this region using a linear fit, and we convolved the background-corrected PSD with a 7 $\mu$Hz wide Gaussian kernel. The final $\nu_{\rm max}$ value corresponds to the frequency of the maximum of the smoothed PSD. To measure the large frequency separation ($\Delta \nu$), we computed an autocorrelation of the PSD convolved with a narrow 0.1 $\mu$Hz Gaussian kernel, and we adopted our final $\Delta \nu$ from the peak closest to the predicted value from \citet{stello2009}. Using this method we obtained the following asteroseismic quantities: $\nu_{\rm max} = 179.69 \pm 4.22$ $\mu$Hz and $\Delta\nu = 14.27 \pm 0.01 $ $\mu$Hz. The uncertainties correspond to the standard deviation of 1000 bootstrap randomization of the data. 
Finally, following the scaling relations presented in \cite{stello2017}, we estimated an asteroseismic mass of 1.22 $\pm$ 0.09 \msun and radius of 4.77 $\pm$ 0.11 \rsun for HIP\,75092. The resulting asteroseismic mass and radius are in good agreement with the spectroscopic values.

\subsection{Lithium abundance and C$^{12}/$\,C$^{13}$ isotopic ratio}

We computed the lithium (Li) abundance for all four stars by using spectral synthesis around the Li doublet at $6708\, \mathrm{\AA}$, under the assumption of local-thermodynamic-equilibrium (LTE). Briefly, the stellar spectra are fitted by a synthetic spectrum which is generated with MOOG \citep[2018 version;][]{moog}, which accounts for possible blends, and allowing some parameters such as continuum position or the stellar radial velocity to slightly change.
For the MOOG input line list, we used a combination of those presented in \citet{Melendez2012} and \citet{Carlberg2012}, and we used the ATLAS9 atmospheric models \citep{CastelliKurucz2004}. A detailed description of this method will be presented in a fourth-coming paper (Aguilera-G\'omez et al. in prep). Unfortunately, we found only upper limits for all four giant stars, which are listed in Table \ref{stellar_par}.

\begin{figure}
       \includegraphics[angle=0,scale=0.39]{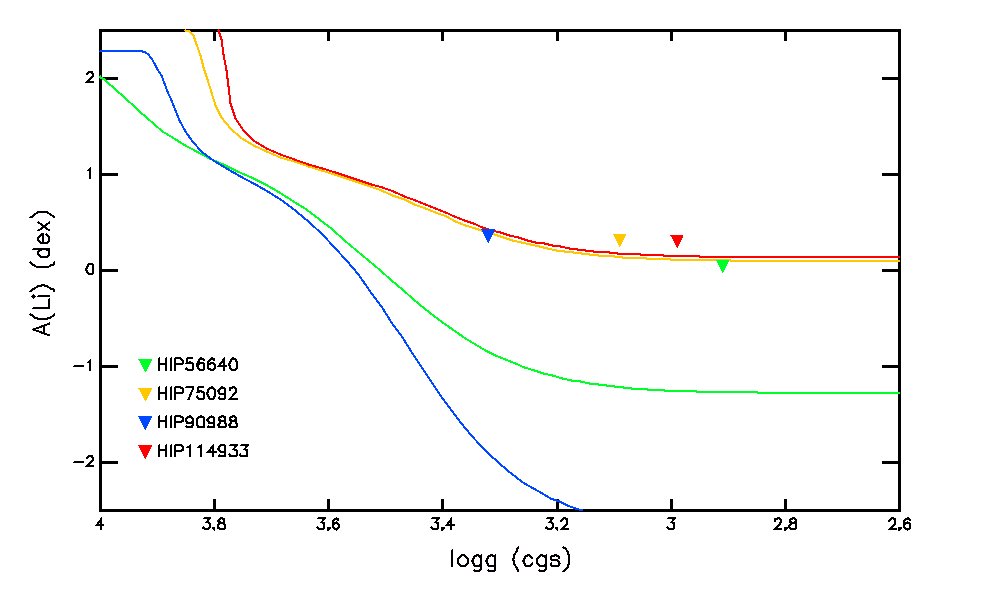}
    \caption{ Theoretical Lithium abundance evolution from the YREC models, for the four giant stars (solid lines). The triangles 
    represent the Lithium abundance upper limits.  \label{fig_Li}}
\end{figure}
These upper limits are consistent with expectations of red giants that have already finished their first dredge-up depletion, when the convective envelope of the star deepens at the beginning of the red giant branch phase (see Figure \ref{hrdiagram}). Figure \ref{fig_Li} shows the Li abundance evolution for these giant stars, calculated using canonical Yale Rotating Evolutionary Code (YREC) models \citep{yrec}. The measured Lithium abundance upper limits are overplotted (filled triangles). As can be seen, the resulting upper limits are above the theoretical values. \newline \indent
Similarly, we computed the carbon isotopic ratio $\mathrm{^{12}C/^{13}C}$, following a similar procedure to that described in \citet{Carlberg2012}. We use spectral synthesis to fit the CN bands (both $\mathrm{^{12}C/N}$ and $\mathrm{^{13}C/N}$) in the region between $8000$ and $8006\,\mathrm{\AA}$. In this spectral region, telluric lines can complicate severely the determination of the ratio. To avoid incorrect measurements, we analyze all of the available FEROS spectra, and we derive the isotopic ratio only in those cases where the telluric lines do not fall near the CN bands. A more detailed description of these measurements will be found in Aguilera-G\'omez et al. (in prep).
We report these values in table \ref{stellar_par}. In the case of HIP\,56640 we could only obtain a lower limit. All of the found carbon isotopic ratios are consistent with first ascending red giant branch stars. The carbon isotopic ratio in these stars is expected to decrease from the solar ratio $\mathrm{^{12}C/^{13}C}\sim90$ to $\mathrm{^{12}C/^{13}C}\sim25$. The values found here confirm the evolutionary phase of the giants, with no extra mixing in their interiors, that would decrease the ratio. Notice that higher values of the carbon isotopic ratio imply a larger error in the measurement. This is because at high ratios, a large difference in $\mathrm{^{13}C}$  only implies very small changes in the spectra, thus making it difficult to identify the best fit for the observations. Even small changes to the continuum placement can affect the ratio. However, in spite of the large errors, the carbon isotopic ratio is large and consistent with normal red giant branch stars going through their first dredge-up dilution.

\section{Orbital fitting}
\label{sec4}

To model the RV data and obtain orbital parameters of the companions we used the {\em The Exo-Striker}
fitting toolbox\footnote{\url{https://github.com/3fon3fonov/exostriker}} \citep{Trifonov2019_es}.  
As a starting point, for each target we performed frequency analysis of the combined RV time series.
The GLS analyses of the available data showed that all four targets contain a very significant periodic
RV signal with semi-amplitudes well above the RV uncertainties.
The period corresponding to the maximum GLS power, phase, and amplitude were 
used as an initial guess of the parameter optimization process, which aims to 
determine the best-fit Keplerian orbital parameters of the planetary candidates.
With {\em The Exo-Striker} we used Simplex algorithm \citep{NelderMead65}, which optimizes 
the negative logarithm of the likelihood function ($-\ln\mathcal{L}$) coupled with a Keplerian model.
The optimized parameters are the RV semi-amplitude $K$, orbital period $P$, eccentricity $e$, 
argument of periastron $\omega$, mean anomalies $M$ for a given epoch (from which we derived 
the time of periastron passage $T_p$), and the RV zero-point offset for each data set.
Additionally, in the case of HIP\,90988, we included a linear ($\dot{\gamma}$) RV trend as an extra fitting parameter.
The parameter uncertainties were estimated using the Markov Chain Monte Carlo (MCMC) sampler \texttt{emcee} \citep{emcee}
included in {\em The Exo-Striker}. We adopted only flat parameter priors and we ran
multiple MCMC chains in parallel starting from the best-fit parameters. 
We adopted the median absolute deviation 
of the MCMC posterior distributions to serve as a $1\sigma$ uncertainty estimate of the parameters.
We note that before searching for the global minimum, we added in 
quadrature 7 \ms to the RV instrumental errors to account for the excess of RV noise (e.g. \citealt{wit2016a}).
We find that this prior jitter level of 7 \ms is well justified keeping in 
mind the known instrumental systematics of FEROS, UCLES and CHIRON, which are of the same order of magnitude.
In addition, evolved stars exhibit short period p-mode oscillations, which contribute to the RV noise. 
For instance, the scaling relation from \citet{kjeldsen1995} suggest 
RV scatter level, that is in a good agreement with the adopted RV scatter in this work.
As a matter of completeness, however, in our modeling scheme we also optimized the RV "jitter" for each data set \citep{Baluev2009}.
We find that the jitter estimates are very small, often consistent with 0 m\,s$^{-1}$,
which means that the prior jitter level of 7 \ms is indeed adequate.
The resulting orbital parameters from our RV analysis and the corresponding uncertainties 
are listed in Table \ref{orb_par}.

\subsection{HIP\,56640 = HD\,100939}

\begin{figure}
       \includegraphics[angle=0,scale=0.45]{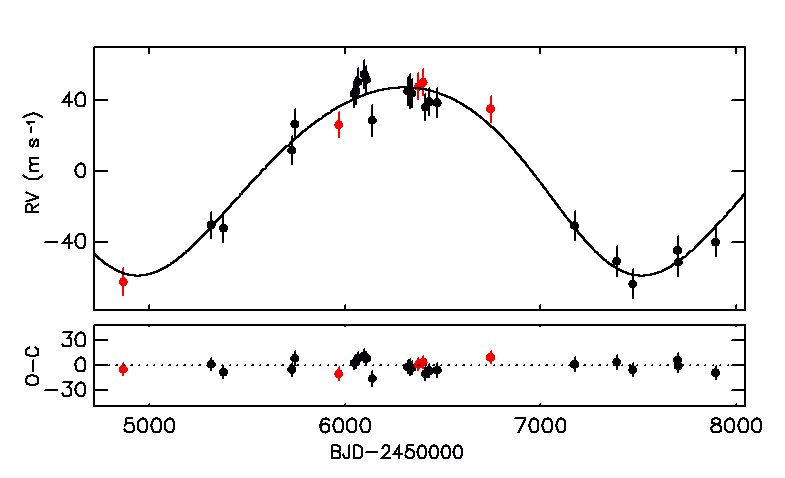}
    \caption{Radial velocity variations of HIP\,56640. The black and red dots correspond to FEROS and UCLES velocities, respectively. The best Keplerian solution is over plotted (black solid line). The post-fit residuals are shown in the lower panel. }
    \label{RV_56640}
\end{figure}

We collected a total of 22 FEROS spectra of HIP\,56640 and 5 more UCLES datapoints, covering more than 3000 days. These combined dataset revealed a relatively large RV signal (peak-to-peak variation $\sim$\,100 \ms), which is consistent with the presence of a relatively massive (3.7 \mjup) planetary companion in a long-period orbit (2575 days). Figure \ref{RV_56640} shows the FEROS and UCLES RVs, and the best keplerian fit to the data. The orbital parameters are listed in Table \ref{orb_par}. 

 \subsection{HIP\,75092 = HD\,136295}
 
\begin{figure}
       \includegraphics[angle=0,scale=0.45]{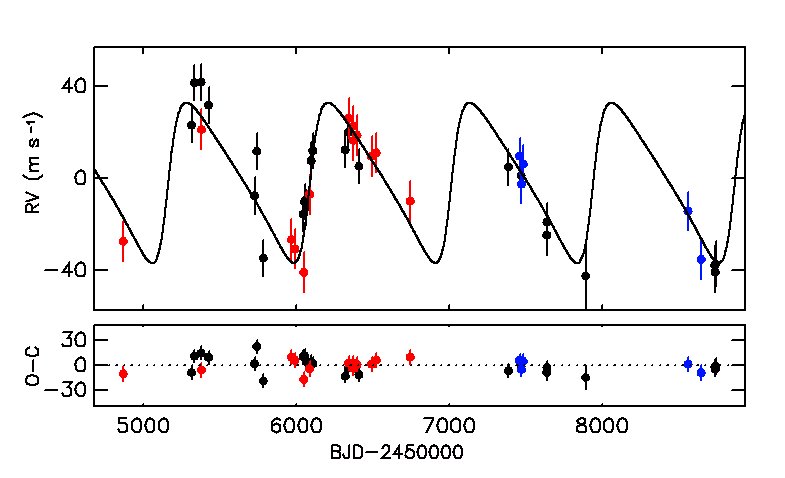}
    \caption{Radial velocity variations of HIP\,75092. The black, red and blue dots correspond to FEROS, UCLES and CHIRON velocities, respectively. The best Keplerian solution is over plotted (black solid line). The post-fit residuals are shown in the lower panel. }
    \label{RV_75092}
\end{figure}

 We observed this star in 23 different epochs with FEROS and we obtained additional 13 UCLES spectra. We computed a GLS periodogram of the combined datasets, and we found a high significance peak (FAP $\sim$ 0.0001) at $\sim$ 940 days. We complemented these datasets, with 5 new CHIRON observations, which confirmed the RV signal detected in the FEROS and UCLES RVs. These combined datasets span more than 10 years of observations.
 Figure \ref{RV_75092} shows the RVs from all three instruments. The best keplerian fit is overplotted. The periodic RV variation observed in this star is consistent with the doppler shift induced by a 1.75 \mjup planet with a period of 926 days. With e=0.42, this is also the most eccentric among the planets presented here, and one of the most eccentric known planets orbiting giant stars. The resulting orbital parameters are listed in Table \ref{orb_par}.
 
 \subsection{HIP\,90988 = HD\,170707}
 
\begin{figure}
       \includegraphics[angle=0,scale=0.36]{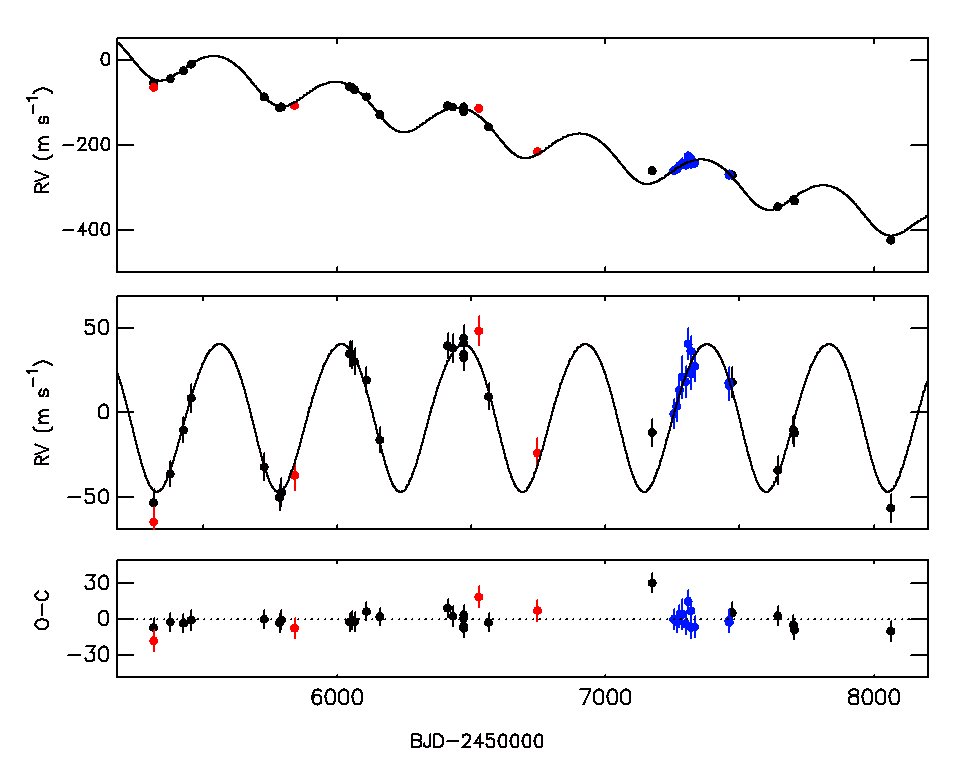}
    \caption{Radial velocity variations of HIP\,90988. The black red and blue dots correspond to FEROS, UCLES and CHIRON velocities, respectively. The Keplerian solution is over plotted (black solid line). The post-fit residuals are shown in the lower panel. }
    \label{RV_90988}
\end{figure}

 We obtained 25 FEROS, 4 UCLES and 13 CHIRON spectra of for HIP\,90988. The combined datasets span more than 2700 days. The FEROS and UCLES data revealed a long linear trend, soon after the first couple of RV measurements were obtained. After collecting additional data, it also became evident that there is a $\sim$ 90 \ms peak-to-peak periodic signal superimposed to a RV linear trend. 
 Figure \ref{RV_90988} shows the RV data of HIP\,90988 (upper panel) and the RVs after subtracting the long-term linear trend (middle panel). The best-fit keplerian solution is overplotted (black solid line). The orbital parameters are listed in Table \ref{orb_par}. The observed RV curve is consistent with the presence of a $\sim$ 1.9 \mjup planet in a 454-day and low-eccentricity orbit (e=0.08).\newline \indent
 Finally, from the linear trend (d$v$/d$t$ = -48.99 $\pm$ 0.83 \ms yr$^{-1}$) we estimated the minimum mass of the outer companion that is compatible with this value. 
 For this we computed numerically different orbital configurations with increasing orbital period and companion mass, under the assumption of m$_c$ $<<$ M$_\star$ and fixing the eccentricity to zero. We then imposed that a synthetic curve matches the data if the difference in the observed and simulated RV acceleration term are consistent at the 1-$\sigma$ level and the residuals of the fit is at the $\sim$ 5 \ms. Here the residuals correspond to the difference between the Keplerian synthetic model and the linear fit to the Keplerian model computed at the time of the observations. We note that a significantly larger value of the residual would be an indication of a quadratic RV term, and it would be present in the real data. Using this method we obtained the following results: m$_c$\,sin$i$ $\gtrsim$ 24 \mjup and $a_c \gtrsim$ 9.0 AU. The  minimum orbital period of $\sim$ 23 years, corresponds to $\sim$ 3 times the observational time span. The resulting minimum mass of the outer companion is well in the planetary to brown dwarf mass regime.  
 Moreover, the minimum angular separation ($\sim$ 95 mas), is very close to inner detection limits for high-contrast imaging instruments such as SPHERE \citep{beuzit2019}, making this system is an interesting candidate for further high-contrast imaging studies to characterize the outer companion.  

\subsection{HIP\,114933 = HD\,219553}

\begin{figure}
       \includegraphics[angle=0,scale=0.45]{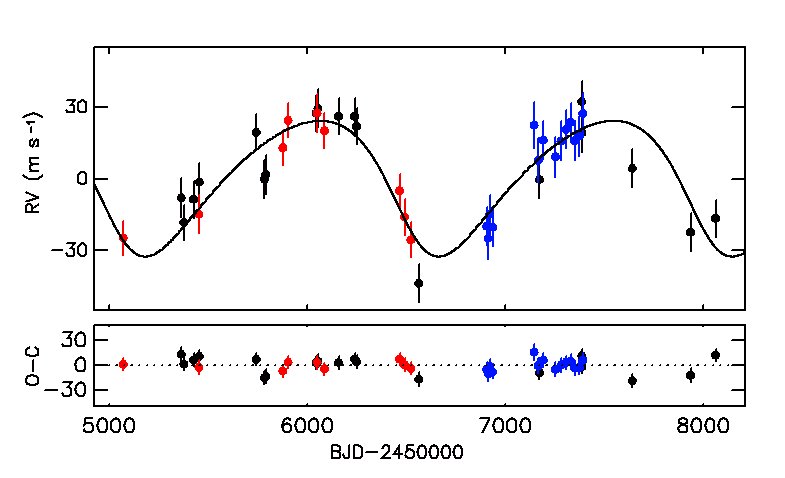}
    \caption{Radial velocity variations of HIP\,114933. The black, red and blue dots correspond to FEROS, UCLES and CHIRON velocities, respectively. The Keplerian solution is over plotted (black solid line). The post-fit residuals are shown in the lower panel. }
    \label{RV_114933}
\end{figure}

We observed HIP\,144933 between August 2009 and November 2017. We collected a total of 19 FEROS, 9 UCLES and 14 CHIRON spectra for this star. The corresponding velocities revealed a long period periodic variation with an amplitude of $\sim$ 30 \ms, most likely induced by a 1.9 \mjup planet orbiting at 2.8 AU from the host star. No evidence of a third body in the system is observed for this star.  Figure \ref{RV_114933} shows the RV variations for all three datasets and the best-keplerian fit. The best-fit orbital elements of \plansix\, and their estimated uncertainties  are listed in Table \ref{orb_par}.
 
\begin{table*}
\centering
\caption{Orbital parameters \label{orb_par}}
\begin{tabular}{lrrrr}
\hline\hline
\vspace{-0.3cm} \\
                             & HIP\,56640\,$b$ & HIP\,75092\,$b$ & HIP\,90988\,$b$  & HIP\,114933\,$b$ \\
\hline \vspace{-0.3cm}                                                                               \\
$K$ [m\,s$^{-1}$]            &   53.3\mm2.0    &   34.8\mm3.9    &  43.6\mm2.1      & 28.4\mm2.5    \\ 
$P$ [day]                    & 2574.9\mm86.1   &  926.4\mm12.8   & 453.9\mm2.0      & 1481.6\mm61.7 \\ 
$e$                          &   0.12\mm0.04   &   0.42\mm0.10   &  0.08\mm0.05     & 0.21\mm0.08  \\
$\omega$ [deg]               &  157.2\mm41.5   &  262.4\mm16.7   & 161.6\mm23.9     & 134.2\mm29.9 \\ 
$a$ [au]                     &   3.73\mm0.08   &   2.02\mm0.02   &  1.26\mm0.01     & 2.84\mm0.08  \\
\mplan [$M_{\rm jup}$]       &   3.67\mm0.14   &   1.79\mm0.16   &  1.96\mm0.09     & 1.94\mm0.17  \\ 
$T_{\rm P}$ -2450000 [day]   & 4813.9\mm347.5  & 4230.5\mm49.0   & 5309.2\mm30.2    & 5059.4\mm104.6 \\ 
$\dot{\gamma}$ [\msyr]       &    $\dots$      &      $\dots$    & -49.0\mm0.8    &  $\dots$ \\
$\gamma_0$ (feros) [\ms]     &  -35.4\mm6.6    &   2.9\mm2.6     & 146.6\mm2.8      & -4.8\mm2.9 \\ 
$\gamma_0$ (ucles) [\ms]     &  -10.0\mm2.2    &  -9.7\mm2.9     & 107.4\mm13.0     & 12.1\mm2.9 \\
$\gamma_0$ (chiron) [\ms]    &    $\dots$      &   7.1\mm 5.9    & 252.6\mm6.0      & -7.0\mm3.4 \\
$\chi_{\nu}^2$               &    1.2          &       1.2       & 1.1              & 1.0   \\
RMS [m\,s$^{-1}$]            &    7.4          &       9.7       & 8.1              & 10.0  \\

\vspace{-0.3cm} \\\hline\hline
\end{tabular}
\end{table*}

\section{Companion detection limits}\label{sec:detectability}

\begin{figure*}
       \includegraphics[angle=0,scale=0.70]{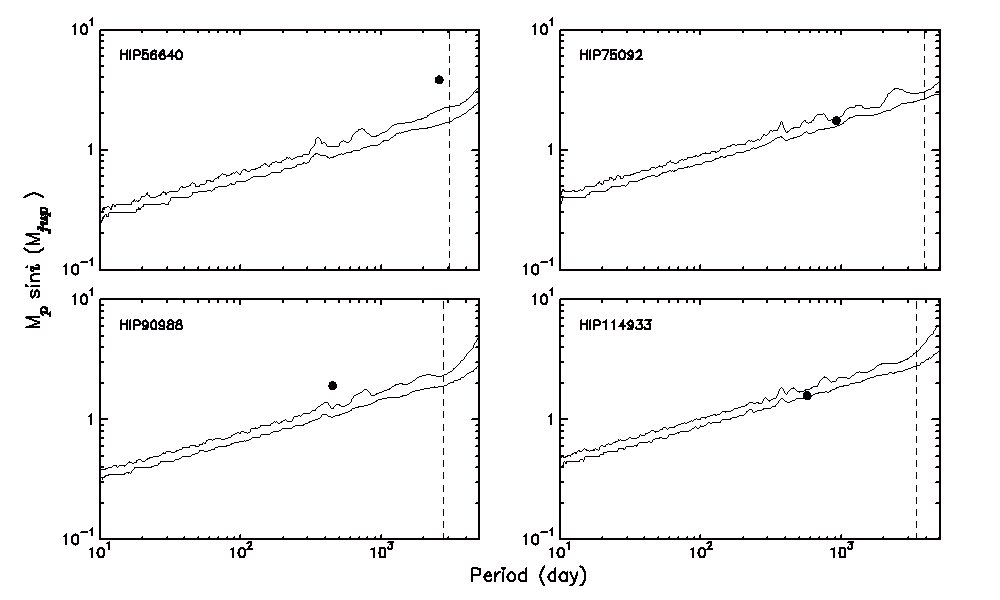}\label{det_planets}
    \caption{Detection limits for all 4 host stars. The solid lines correspond to 50\% and 95\% detection efficiency (from bottom to top). The black dots show the position of the companion. The vertical dashed lines corresponds to the total RV time coverage.}
    \label{fig:detectability}
\end{figure*}

We computed detection limits for all four stars, to assess which planets we are able to detect with these RV datasets. For this, we used the RMS approach, following previous works (e.g. \citealt{bowler2010,wittenmyer2020b}).   
Briefly, we computed a large number of synthetic RVs at each observing epoch, corresponding to different companion mass and orbital period, and zero eccentricity. We then compared the observed RMS of the synthetic datasets to the observed post-fit residuals. We imposed that a system is detected if the RMS of the synthetic RVs is 2.5 times larger than the post-fit residuals RMS of the real data. The results are presented in Figure \ref{fig:detectability}. The solid lines correspond to 50\% and 95\% detectability curves (from bottom to top, respectively). The position of the planetary companions (black dots) and total observing timespan (vertical dashed lines) are overplotted. It can be seen that we are very sensitive ($\gtrsim$ 95\% completeness) to planets with M$_P$ $\gtrsim 1.0$ \mjup and orbital period $P \lesssim$ 100 days, or M$_P$ $\gtrsim 2.0$ \mplan up to periods of $\sim$ 1000 days. 

\section{Stellar activity analysis \label{sec:stellar_activity}}

To investigate if stellar phenomena are responsible for the observed RV variations (e.g. \citealt{boisse11}), we analyzed in detail the \hipparcos and the All Sky Automated Survey (ASAS; \citealt{asas}) $V$-band photometry of all four giant stars. In the analysis we only included \hipparcos data with quality flag equal to 0 and 1. Similarly, we only used A and B grade ASAS data. For both datasets, we applied a 3\,$\sigma$ clipping filter. We then combined the two datasets, after applying a zero-point offset between them, and we computed a Lomb-Scargle periodograms to the combined data. This procedure not only allows us to search for long-period signals exceeding their individual total timespan, but also it boosts the significance of periodic signals present in the two individual datasets.
We found no significant peaks in the periodograms in all four stars. \newline \indent
Similarly, we computed the stellar activity S-indexes from the FEROS data, following the procedure described in \citet{jones2017}. Again, we found no significant correlation between the S-index and the observed RVs. 
The Hipparcos magnitude (H$_p$), S-index and their corresponding variability\footnote{The variability corresponds to the standard deviation from all individual measurements.} are listed in Table \ref{activity_indicators}. As can be seen these stars are photometrically and chromospherically very stable. This is not surprising since they were selected against photometric variability. \newline \indent
Finally, to discard line asymmetry variations as the cause of the observed RVs, we
computed FEROS bisectors following the method presented in \citet{jones2017}. We then compared the BVS values with the observed radial velocities. In the case of HIP\,90988 we first subtracted the RV linear trend. 
We found no significant correlation between the BVS and the corrected RVs, except for HIP\,56640, that present a moderate level of correlation ($\rho$ = 0.43 $\pm$ 0.18). However, this correlation is probably explained by known instrumental profile variations with time \citep{jones2017}. 


\begin{table}
\centering
\caption{Stellar activity indicators \label{activity_indicators}}
\begin{tabular}{lrrrr}
\hline\hline
\vspace{-0.3cm} \\
HIP     &  H$_p$ & $\sigma_H$ & S-index & $\sigma_S$\\
\hline \vspace{-0.3cm} \\
56640   & 8.11   & 0.010      &   0.133  &  0.005   \\  
75092   & 7.28   & 0.009      &   0.133  &  0.006   \\ 
90988   & 7.92   & 0.014      &   0.133  &  0.007   \\ 
114933  & 7.42   & 0.011      &   0.128  &  0.006   \\  
\vspace{-0.3cm} \\\hline\hline
\end{tabular}
\end{table}
                             
\section{Hipparcos Astrometry}\label{sec:astrometry}

To discard that the RV signals are induced by low inclination binary companions, we attempted to measure the orbital inclination angles using the combined \hipparcos astrometric data and the available RV measurements, following the procedure described in \citet{SAH11} and \citet{jones2017}. Using this method, we have derived inclination angles for three of the binary companions detected by the EXPRESS program (Jones et al. in preparation). In this particular case, we could not detect the astrometric signal induced by the companion. Given the planet-to-star mass ratio, the expected astrometric amplitude signals are expected to be $\lesssim$ 0.1 miliarcsec, which is  well below the \hipparcos precision. 
Moreover, in all four cases, the \gaia DR2 astrometric excess noise is equal to zero, meaning that there is no indication of an astrometric motion induced by a companion. In fact, by assuming edge-on orbits, the astrometric signals for four six stars is at the $\sim$ 20-30 $\mu$\,arcsec level, which is comparable to the \gaia detection limit. 

\section{Discussion}\label{sec:conclusions}

\subsection{Summary}
In this paper we present precision radial velocities of four giant stars that have been targeted for about a decade by the EXPRESS and PPPS radial velocity surveys. The radial velocities have been computed from spectroscopic data collected with three different instruments mounted on 1-3 meter-class telescopes in the Southern Hemisphere. The RV data have revealed periodic signals which are most likely explained by the presence of substellar companions. From the Keplerian fits we derived companion minimum masses between $\sim$ 1.8 and 3.7 \mjup, orbital periods of 1.2\,-\,7.1 years and eccentricities between 0.08 and 0.42. 
Additionally, HIP\,90988 presents a long-term RV trend of -49.0 \ms \,yr$^{-1}$, which is most likely induced by an outer body with a minimum mass in the planetary mass regime (m$_c$\,sin$i$ $\gtrsim$ 24 \mjup). We note that either further RV follow-up or high-contrast imaging observations are needed to confirm or discard the planetary nature of the companion.  \newline \indent
From our combined photometric and spectroscopic analysis, we found that all four host stars are low-luminosity giants with masses between 1.04 and 1.39 \msun. Based on their position on the HR diagram, these stars can be unambiguously identified as first ascending RGB stars, and have completed their first dredge-up (see Figure \ref{hrdiagram}). Moreover, given their relatively small radii ($\lesssim$ 0.025 AU), they are not expected yet to have engulfed planets with $a \gtrsim$ 0.05\,-\,0.1 AU \citep[e.g.][]{villaver2009,kunitomo2011}. 

\subsection{Occurrence rate from the EXPRESS and PPPS common targets}

\begin{figure}
       \includegraphics[angle=0,scale=0.35]{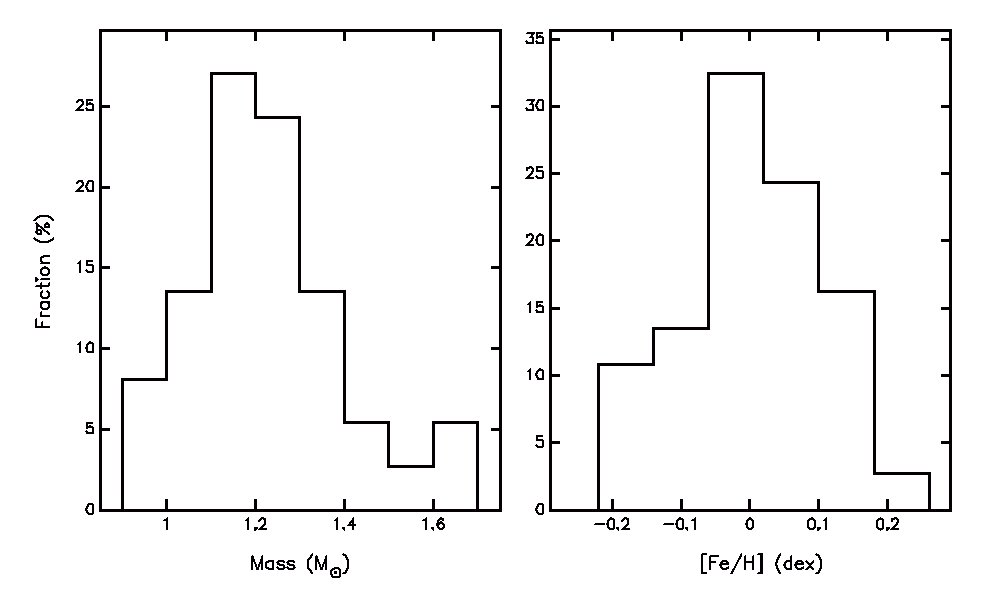}
    \caption{Mass and metallicity distribution of the 37 targets in common between the EXPRESS and PPPS.}
    \label{fig:mass_metallicity}
\end{figure}

\begin{figure}
       \includegraphics[angle=0,scale=0.38]{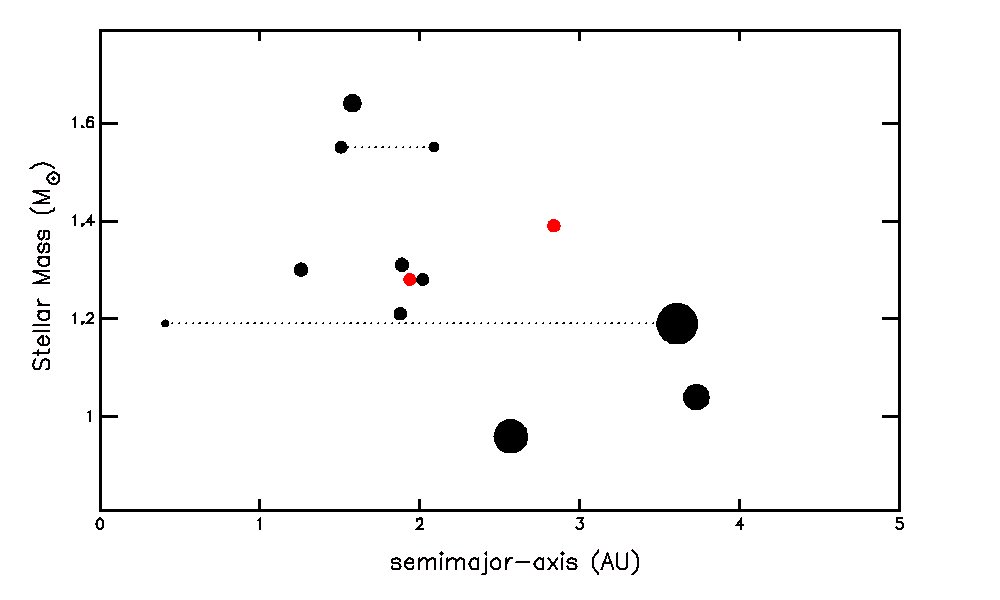}
    \caption{Semimajor axis distribution versus the mass of the host star, for the common planets between the EXPRESS and PPPS. The black filled circles correspond to the confirmed and published planets, while the red filled circles correspond to the unpublished planet candidates. The symbol size is proportional to the minimum mass of the planet. The dotted lines connect the position of the inner and outer planets in multiple systems.}
    \label{fig:semimajor_ms}
\end{figure}

Figure \ref{hrdiagram} shows the position in the HR diagram for all the 37 stars in common between the EXPRESS and PPPS. Similarly, Figure \ref{fig:mass_metallicity} shows their stellar mass and metallicity distribution\footnote{The stellar mass and metallicty of the 37 stars was derived using the newest version of SPECIES. See section \ref{sec3}}. The median mass and metallicity of this sample is 1.21 $\pm$ 0.16 \msun and 0.0 $\pm$ 0.1 dex, respectively. Interestingly, 10 out of the 37 stars have been identified as spectroscopic binaries, and four of them in a compact system ($a \lesssim$ 
5.0 AU; \citealt{bluhm2016}), where the formation and survival of circumprimary (s-type) giant planets is fully suppressed (see \citealt{moe2019} and references therein).
For each star, we have collected $\gtrsim$ 20 individual spectra between 2009 and 2019. For the planet candidate host stars we have taken additional spectra, adding-up typically $\gtrsim$ 40 RV epochs. This combined dataset have allowed us to efficiently detect giant planets (\mplan $\gtrsim$ 1-2 \mjup) orbiting their parent star up to an orbital distance of $\sim$ 5 AU (see Figure \ref{fig:detectability}). In fact, including the results presented here, we have found a total of 13 giant planets around 11 stars within 5 AU, including two multiple systems \citep{jones2015a,jones2015b,wit2016a}. Figure \ref{fig:semimajor_ms} shows the semimajor axis distribution versus the stellar mass for all 13 planets detected around the 37 stars. The position of the two unpublished planet candidates is also shown. As can be seen, the orbital distribution is substantially different when compared to giant planets orbiting solar-type stars, and is distinctively characterized by a desert of close-in planets ($a \lesssim 0.5 $ AU) and a complete absence of Hot Jupiters \citep{sato2008,dollinger2009}. We note that the innermost planet ($a \sim$ 0.4 AU) among this sample is found in a multiple-system. Considering the high mass ration between the inner and outer planets ($q$ $\sim$ 5) we might speculate that the presence of the outer companion has played an important role in the system orbital configuration. However, the mechanism is not clear, particularly considering that more massive planets can destabilize the orbits of smaller planets \citep{ida2013}.
If we exclude from the analysis the four aforementioned compact binaries, we obtain a surprisingly high fraction of giant stars hosting at least one Jovian planet within 5 AU of $f = 33.3^{+9.0}_{-7.1} \%$. The lower and upper error bars correspond to the 1-$\sigma$ equal-tailed confidence limit, following the Bayesian approach presented in \citep{cameron2011}. 
By restricting the orbital distance to 2.5 and 3.0 AU, we obtained a fraction of $24.2 ^{+8.8}_{-5.9} \%$ and $30.3^{+9.0}_{-6.7} \%$, respectively. 
The former value is significantly higher than the 4.2 $\pm$ 0.7 \% and 8.9 $\pm$ 2.9\% fraction of Jovian planets within 2.5 AU for solar-type stars and higher-mass subgiants, reported by \citet{johnson2007b}. However, our results are compatible at the 1-$\sigma$ level with $f = 26.0 ^{+9.0}_{-8.0}$ within 3 AU, obtained by \citet{bowler2010} from a uniform sample of intermediate-mass subgiants. The median mass and metallicity of the 11 host stars is 1.29 $\pm$ 0.20 and 0.06 $\pm$ 0.11 dex, respectively. This means that the host stars are significantly more massive than the Sun, which probably explains the striking differences in the fraction and orbital distribution when compared to planet orbiting solar-type stars, which is discussed in the next section.

\subsection{Detection fraction and orbital distribution in the context of the core-accretion model}

There are two main ingredients for the formation of giant planets in the core-accretion model, namely the dust and gas content in the protoplanetary disk. A high surface density of dust ($\sigma_d$) is needed to rapidly form planetesimals before the depletion of the dusty component in the disk ($\lesssim$ 3-6 Myr; e.g. \citealt{haisch2001}). 
They will subsequently form protoplanet cores by coagulation and subsequent runaway and oligarchic growth (e.g. \citealt{kokubo1998}). In the inner region of the disk, $\sigma_d$ might not be high enough to form massive cores, but beyond the snow-line this value is predicted to increase by a factor $\sim$ 3-4, leading to $\sim$ 5 times more massive protoplanets \citep{kennedy2008}.
After reaching a mass of $\gtrsim$ 5-10 M$_\oplus$, these cores can efficiently accrete the surrounding gas in the disk (e.g. \citealt{pollack1996}), eventually becoming a gas giant planet. During the gas accretion process onto the protoplanetary core, a significant amount of gas (and a correspondingly high gas surface density; $\sigma_g$), must be present in the disk, before the gas dissipation by accretion in the central star \citep{muzerolle2005}  and photoevaporation \citep{kennedy2009}. 
Having these formation processes in mind, we can understand, at least in a qualitative way, the observed properties of gas giants among the EXPRESS and PPPS common sample as follows. The host stars are slightly metal-rich (median metallicity of +0.06 $\pm$ 0.04 dex), ensuring a relatively high $\sigma_d$ to efficiently form relatively massive protoplanets. Similarly, since the disk mass scales with the stellar mass (e.g. \citealt{andrews2013}), both $\sigma_d$ and $\sigma_g$ are enhanced compared to lower mass stars (e.g. \citealt{laughlin2004}), and thus the minimum core mass to undergo runaway gas accretion can be quickly formed in a wide range of orbital separation \citep{Ida2005}. These protoplanetary cores will then efficiently accrete gas from the disk, resulting in a high fraction of giant planets. In fact, theoretical models predict an increase in the giant planets fraction with the stellar mass, from $\sim$ 1\% at 0.4 \msun, to $\sim$ 10\% at 1.5 \msun \citep{kennedy2008}. However, we found a substantially higher fraction of gas giants at $\sim$ 1.3 \msun, meaning that their formation efficiency might be larger than what is predicted. \newline \indent 
Finally, the larger orbital separation observed for gas giants orbiting intermediate-mass stars when compared to solar-type host stars \citep{johnson2007a} might be explained by the effect of the stellar mass on the disk properties and evolution. In particular, not only can massive enough cores be formed at wider orbital separations, but also the position of the snow line moves outward due to the increasing stellar luminosity with increasing mass. In these two cases, in-situ formation either inside or beyond the snow line occur at larger $a$. Similarly, due to the increasing accretion and photoevaporation rates with increasing stellar mass \citep{muzerolle2005,burkert2007}, the inner part of the disk is also more rapidly depleted (e.g. \citealt{currie2007}). As a result, and due to the shorter disk lifetime, Type-II migration \citep{papaloizou1984} might be rapidly halted, preventing giant planets to move inward, which would explain the relatively large orbital distance and the lack of Hot Jupiters orbiting evolved intermediate-mass stars, as is the case of the planets presented here.

\begin{acknowledgements}

We acknowledge the traditional owners of the land on which the AAT stands, the Gamilaraay people, and pay our respects to elders past and present. C.A.G. acknowledges support from the National Agency for Research and Development (ANID) FONDECYT Postdoctoral Fellowship 2018 Project 3180668. JSJ acknowledges support by FONDECYT grant 1201371, and partial support from CONICYT project Basal AFB-170002. MRD is supported by CONICYT-PFCHA/Doctorado Nacional- 21140646, Chile

\end{acknowledgements}


\begin{appendix} 

\section{Radial velocity tables.}

\begin{table}
\centering
\caption{Radial velocities for HIP\,56640 \label{hip56640_rv}}
\begin{tabular}{lrrr}
\hline\hline
\vspace{-0.3cm} \\
BJD-2450000 & RV (m/s) & err (\ms) & instrument\\
\hline \vspace{-0.3cm}                                                                                                                                 \\
4868.10568  &     -97.9  &  2.5  & UCLES \\
5969.15781  &      -9.0  &  2.1  & UCLES \\
6376.00168  &      12.7  &  2.3  & UCLES  \\
6399.99654  &      15.0  &  2.2  & UCLES  \\
6745.08010  &       0.0  &  2.3  & UCLES  \\
5317.47895   &  -40.0  &  2.3  & FEROS \\
5379.51954   &  -40.6  &  3.2  & FEROS \\
5729.52784   &    1.3  &  3.5  & FEROS \\
5744.48682   &   16.6  &  4.3  & FEROS \\
6047.50514   &   33.2  &  1.9  & FEROS \\
6056.49750   &   36.9  &  3.0  & FEROS \\
6066.52464   &   40.0  &  2.6  & FEROS \\
6099.49205   &   45.9  &  3.8  & FEROS \\
6110.46849   &   39.9  &  2.8  & FEROS \\
6140.50509   &   18.4  &  5.1  & FEROS \\
6321.67847   &   36.8  &  3.7  & FEROS \\
6331.66872   &   23.6  & 10.8  & FEROS \\
6342.63228   &   32.8  &  3.7  & FEROS \\
6412.52140   &   29.6  &  3.0  & FEROS \\
6431.55134   &   32.3  &  3.9  & FEROS \\
6472.50577   &   29.3  &  3.6  & FEROS \\
7174.45993   &  -45.0  &  2.9  & FEROS \\
7389.70522   &  -60.1  &  3.9  & FEROS \\
7471.51094   &  -72.8  &  3.5  & FEROS \\
7700.86516   &  -57.6  &  4.5  & FEROS \\
7703.84316   &  -50.5  &  3.4  & FEROS \\
7895.48289   &  -49.1  &  3.8  & FEROS \\
\vspace{-0.3cm} \\\hline\hline
\end{tabular}
\end{table}

\begin{table}
\centering
\caption{Radial velocities for HIP\,75092 \label{hip75092_rv}}
\begin{tabular}{lrrr}
\hline\hline
\vspace{-0.3cm} \\
BJD-2450000 & RV (m/s) & err (\ms) & instrument\\
\hline \vspace{-0.3cm}                                                                                                                                 \\
4871.26984  & -37.0 & 1.8 & UCLES \\
5382.04745  &  11.7 & 1.7 & UCLES \\
5970.26468  & -36.2 & 1.6 & UCLES \\
5994.18373  & -40.3 & 2.1 & UCLES \\
6052.16806  & -50.5 & 4.1 & UCLES \\
6089.03434  & -16.5 & 1.8 & UCLES \\
6344.25135  &  16.5 & 1.9 & UCLES \\
6375.20140  &   6.9 & 2.1 & UCLES \\
6376.22902  &  13.1 & 1.7 & UCLES \\
6400.07957  &   9.1 & 1.9 & UCLES \\
6494.94664  &   0.0 & 1.8 & UCLES \\
6525.90993  &   1.6 & 1.9 & UCLES \\
6747.15234  & -19.5 & 1.6 & UCLES \\
5317.72681  &    25.4  &  3.1  & FEROS \\
5336.81996  &    41.5  &  3.0  & FEROS \\
5379.72291  &    48.3  &  4.0  & FEROS \\
5428.58902  &    33.7  &  3.5  & FEROS \\
5729.75094  &    -4.4  &  4.1  & FEROS \\
5744.68813  &    17.7  &  3.3  & FEROS \\
5786.60282  &   -36.7  &  3.7  & FEROS \\
6047.70569  &   -14.2  &  2.7  & FEROS \\
6056.70855  &    -4.8  &  3.5  & FEROS \\
6066.70623  &    -8.7  &  3.2  & FEROS \\
6099.29623  &    13.6  &  3.7  & FEROS \\
6110.26591  &    14.3  &  2.8  & FEROS \\
6321.88760  &    16.8  &  2.2  & FEROS \\
6342.90834  &    19.5  &  2.4  & FEROS \\
6412.78446  &    10.7  &  2.9  & FEROS \\
7389.85755  &     4.1  &  3.7  & FEROS \\
7472.90762  &     4.1  &  4.1  & FEROS \\
7641.50993  &   -28.1  &  3.4  & FEROS \\
7642.48597  &    -6.5  &  4.1  & FEROS \\
7895.76138  &   -36.3  & 10.0  & FEROS \\
8739.48670  &   -38.5  &  4.8  & FEROS \\
8742.48605  &   -37.5  &  5.0  & FEROS \\
8745.47800  &   -33.7  &  6.6  & FEROS \\
7462.77437   &   16.9   &  3.7 & CHIRON \\
7474.69989   &     4.7  &  4.1 & CHIRON \\
7486.65653   &    13.4  &  4.1 & CHIRON \\
8564.80694   &    -7.0  &  4.6 & CHIRON \\
8650.71535   &   -28.0  &  4.9 & CHIRON \\
\vspace{-0.3cm} \\\hline\hline
\end{tabular}
\end{table}

\begin{table}
\centering
\caption{Radial velocities for HIP\,90988 \label{hip90988_rv}}
\begin{tabular}{lrrr}
\hline\hline
\vspace{-0.3cm} \\
BJD-2450000 & RV (m/s) & err (\ms) & instrument \\
\hline \vspace{-0.3cm}  \\
5318.31103  &   42.8 &  1.1 & UCLES \\
5842.87753  &    0.0 &  1.9 & UCLES \\
6529.00087  &   -6.7 &  2.2 & UCLES \\
6747.25491  & -108.0 &  1.6 & UCLES \\
5317.79046  &   101.1  &  4.8   & FEROS \\
5379.81633  &   102.8  &  3.1   & FEROS \\
5428.69298  &   122.8  &  2.5   & FEROS \\
5457.60922  &   137.3  &  3.6   & FEROS \\
5729.86849  &    60.2  &  3.7   & FEROS \\
5786.72211  &    29.9  &  3.1   & FEROS \\
5793.73316  &    39.1  &  4.3   & FEROS \\
6047.80409  &    83.5  &  2.4   & FEROS \\
6056.75957  &    85.3  &  3.3   & FEROS \\
6066.77330  &    73.4  &  3.1   & FEROS \\
6110.72484  &    62.4  &  2.9   & FEROS \\
6160.67589  &    16.7  &  2.5   & FEROS \\
6412.68341  &    42.2  &  3.0   & FEROS \\
6431.81361  &    33.0  &  3.7   & FEROS \\
6472.72368  &    34.3  &  2.5   & FEROS \\
6472.77298  &    26.6  &  3.0   & FEROS \\
6472.81158  &    21.1  &  3.3   & FEROS \\
6472.83915  &    37.3  &  4.0   & FEROS \\
6565.55337  &   -14.8  &  3.2   & FEROS \\
7174.65398  &  -121.6  &  3.7   & FEROS \\
7472.91065  &  -123.9  &  5.8   & FEROS \\
7642.49794  &  -197.4  &  3.9   & FEROS \\
7700.51469  &  -179.9  &  3.9   & FEROS \\
7704.50942  &  -194.0  &  4.2   & FEROS \\
8063.50122  &  -277.7  &  3.8   & FEROS \\
  7255.55940 &  -7.7 &  4.5 & CHIRON \\
  7266.50506 &  -4.8 &  5.4 & CHIRON \\
  7276.58578 &   3.4 &  6.0 & CHIRON \\
  7287.54225 &   9.4 & 10.2 & CHIRON \\
  7299.56882 &   5.3 &  5.5 & CHIRON \\
  7308.52898 &  26.4 &  6.0 & CHIRON \\
  7318.54706 &  20.9 &  5.1 & CHIRON \\
  7319.53890 &   7.3 &  5.9 & CHIRON \\
  7332.51883 &   9.9 &  5.4 & CHIRON \\
  7460.88641 & -16.8 &  5.8 & CHIRON \\
  7460.89703 & -17.5 &  4.7 & CHIRON \\
  7460.90765 & -16.8 &  5.4 & CHIRON \\
  7461.91885 & -18.9 &  4.4 & CHIRON \\
\vspace{-0.3cm} \\\hline\hline
\end{tabular}
\end{table}

\begin{table}
\centering
\caption{Radial velocities for HIP\,114933 \label{hip114933_rv}}
\begin{tabular}{lrrr}
\hline\hline
\vspace{-0.3cm} \\
BJD-2450000 & RV (m/s) & err (\ms) & instrument \\
\hline \vspace{-0.3cm}  \\
5074.27623  & -12.6  &  1.6  & UCLES \\
5455.99452  &  -2.7  &  3.9  & UCLES \\
5879.94508  &  25.1  &  1.9  & UCLES \\
5905.94188  &  36.7  &  1.6  & UCLES \\
6052.30456  &  39.6  &  3.6  & UCLES \\
6089.29296  &  32.2  &  1.9  & UCLES \\
6469.28447  &   7.1  &  1.6  & UCLES \\
6494.23969  &  -3.8  &  2.5  & UCLES \\
6526.06839  & -13.4  &  1.8  & UCLES \\
5366.95051  &   -12.6  &  4.4 & FEROS \\
5379.92813  &   -23.0  &  2.4 & FEROS \\
5428.81114  &   -13.3  &  3.3 & FEROS \\
5457.71609  &    -6.1  &  3.5 & FEROS \\
5744.87878  &    14.7  &  2.4 & FEROS \\
5786.88660  &    -4.8  &  3.9  & FEROS \\
5793.88608  &    -2.9  &  4.2  & FEROS \\
6047.92006  &    22.8  &  2.9  & FEROS \\
6056.85229  &    24.8  &  3.5  & FEROS \\
6160.79552  &    21.4  &  2.9  & FEROS \\
6241.61801  &    21.4  &  2.5  & FEROS \\
6251.64467  &    17.2  &  3.0  & FEROS \\
6565.67782  &   -48.5  &  3.7  & FEROS \\
7174.85230  &    -5.0  &  3.1  & FEROS \\
7388.54810  &    27.6  &  4.5  & FEROS \\
7389.54623  &    14.6  &  4.1  & FEROS \\
7643.76170  &    -0.4  &  3.7  & FEROS \\
7937.94079  &   -27.1  &  4.0  & FEROS \\
8064.52764  &   -21.3  &  3.3  & FEROS \\
6908.67217 &  -26.8  &  3.8 & CHIRON \\
6916.59716 &  -31.8  &  5.2 & CHIRON \\
6924.61406 &  -21.7  &  3.7 & CHIRON \\
6938.59670 &  -27.2  &  4.0 & CHIRON \\
7147.91066 &   15.5  &  6.8 & CHIRON \\
7168.85473 &    0.8  &  5.9 & CHIRON \\
7193.77769 &    9.3  &  3.9 & CHIRON \\
7255.73338 &    2.3  &  4.6 & CHIRON \\
7283.69092 &    8.9  &  4.1 & CHIRON \\
7308.60370 &   13.7  &  3.7 & CHIRON \\
7332.58470 &   16.7  &  4.1 & CHIRON \\
7353.54492 &    9.0  &  4.2 & CHIRON \\
7376.53217 &   11.0  &  4.5 & CHIRON \\
7394.52487 &   20.4  &  4.6 & CHIRON \\
\vspace{-0.3cm} \\\hline\hline
\end{tabular}
\end{table}

\end{appendix}

\begin{thebibliography}{}


\bibitem[Andrews et al. (2013)]{andrews2013} Andrews, S. M., Rosenfeld, K. A., Kraus, A. L., \& Wilner, D. 2013, ApJ, 771,129

\bibitem[Baluev(2009)]{Baluev2009} Baluev, R. V. 2009, MNRAS, 393, 969
\bibitem[Baranne et al.(1996)]{baranne1996} Baranne, A., Queloz, D., Mayor, M. et al. 1996, \aap, 119, 373
\bibitem[Barbato et al.(2018)]{barbato2018} Barbato, D., Bonomo, A. S., Sozzetti, A. \& Morbidelli, R. 2018, arXiv:1811.08249
\bibitem[Bergmann et al.(2020)]{bergmann2020} Bergmann, C., Jones, M. I., Zhao, J. et al. 2020, accepted in PASA
\bibitem[Beuzit et al.(2019)]{beuzit2019} Beuzit, J.~L., Vigan, A, Mouillet, D., et al. 2019, A\&A, 631, 155
\bibitem[Bluhm et al.(2016)]{bluhm2016} Bluhm, P., Jones, M. I., Vanzi, L. et al. 2016, A\&A, 593, 133
\bibitem[Boisse et al.(2011)]{boisse11} Boisse, I., Bouchy, F., Hebrard, G. et al. 2011, A\&A, 528, 4
\bibitem[Borucki et al.(2010)]{kepler} Borucki, W. J., Koch, D., Basri, G., et al. 2010, Science, 327, 977
\bibitem[Bovy et al.(2016)]{bovy2016} Bovy, J., Rix, H.-W.,  Green, G. M., Schlafly, E. F. \& Finkbeiner, D. P. 2016, ApJ, 818, 130
\bibitem[Bowler et al.(2010)]{bowler2010} Bowler, B. P., Johnson, J. A., Marcy, G. W. et al. 2010, ApJ, 709, 396
\bibitem[Brahm et al.(2017)]{brahm2017} Brahm, R., Jord\'an, A. \& Espinoza, N. 2016, PASP, 129, 34002
\bibitem[Bressan et al.(2012)]{bressan2012} Bressan A., Marigo P., Girardi L. et al. 2012, MNRAS, 427, 127
\bibitem[Brown et al.(2018)]{gaia2018} Brown, A. G. A., Vallenari, A., Prusti, T. et al. 2018, A\&A, 616, 1
\bibitem[Burkert \& Ida(2007)]{burkert2007} Burkert, A., \& Ida, S. 2007, ApJ, 660, 845
\bibitem[Butler et al.(1996)]{butler1996} Butler, R. P., Marcy, G. W., Williams, E. et al. 1996, PASP, 108, 500

\bibitem[Cameron(2011)]{cameron2011} Cameron, E. 2011, PASA, 28, 128
\bibitem[Castelli \& Kurucz(2004)]{CastelliKurucz2004} Castelli, F. \& Kurucz, R.~L. 2004, ArXiv e-prints [astro-ph/0405087]
\bibitem[Carlberg et al.(2012)]{Carlberg2012} Carlberg, J.~K., Cunha, K., Smith, V. \& Majewski, S.~R. 2012, ApJ, 757, 109
\bibitem[Currie et al.(2007)]{currie2007} Currie, T., Balog, Z., Kenyon, S. J. et al. 2007, ApJ, 659, 599
\bibitem[Cutri et al.(2003)]{cutri2003} Cutri, R. M., Skrutskie, M. F., van Dyk, S., et al. 2003,
VizieR Online Data Catalog, 2246, 0
\bibitem[Diego et al.(1990)]{diego90} Diego, F., Charalambous, A., Fish, A.~C., \& Walker, D.~D.\ 1990, \procspie, 1235, 562 
\bibitem[Dohlen et al.(2008)]{dohlen2008} Dohlen, K., Langlois, M., Saisse, M., et al. 2008, SPIE Conf. Ser., 7014, 70143L
\bibitem[D\"{o}llinger et al.(2009)]{dollinger2009} D\"{o}llinger, M.P., Hatzes, A. P., Pasquini, L., Guenther, E. W. \& Hartmann, M. 2009, A\&A 505, 1311
\bibitem[Dotter(2016)]{Dotter2016} Dotter, A. 2016, ApJS, 222, 8
\bibitem[Duncan et al.(1991)]{duncan1991} Duncan, D. K., Vaughan, A. H., Wilson, O. C., et al. 1991, ApJS, 76, 383

\bibitem[Feroz et al.(2009)]{Feroz2009} Feroz, F., Hobson, M. P. \& Bridges, M. 2009, MNRAS, 398, 1601
\bibitem[Foreman-Mackey et al.(2013)]{emcee} Foreman-Mackey, D., Hogg, D. W.,  Lang, D. \& Goodman, J. 2013, PASP, 125, 306
\bibitem[Frink et al.(2001)]{frink2001} Frink, S., Quirrenbach, A., Fischer, D., R\"{o}ser, S. \& Schilbach, E. 2001, PASP, 113, 173
\bibitem[Gaia collaboration et al.(2018)]{gaiadr2} Gaia collaboration, Brown, A.~G.~A., Vellenari, A., et al. 2018, ArXiv e-prints, arXiv:1804.09365

\bibitem[Haisch et al.(2001)]{haisch2001} Haisch, E. Jr., Lada, E. A. \& Lada, C. J. 2001, ApJ, 553, 153
\bibitem[Han et al.(2010)]{han2010} Han, I., Lee, B. C., Kim, K. M. et al. 2010, A\&A, 509, 24
\bibitem[Hardegree-Ullman et al.(2019)]{hardegree2019} Hardegree-Ullman, K. K., Cushing, M. C., Muirhead, P. S. \& Christiansen, J. L. 2019, AJ, 158, 75
\bibitem[Hatzes et al.(2005)]{hatzes2005} Hatzes, A. P., Guenther, E. W., Endl, M. et al. 2005, A\&A, 437, 743
\bibitem[H\o g et al.(2000)]{hog2000} H\o g, E., Fabricius, C., Makarov, V. et al. 2000, A\&A, 355, 27
\bibitem[Houk(1978)]{houk1978} Houk N., Smith-Moore M., 1988, Michigan Catalogue of Two-dimensional Spectral Types for the HD Stars, Vol. 2, Declinations -53.0$^{\circ}$ to -40.0$^{\circ}$. Department of Astronomy, University of Michigan, Ann Arbor, MI
\bibitem[Houk(1982)]{houk1982} Houk N. 1982, Michigan Catalogue of Two-dimensional Spectral Types for the HD Stars, Vol. 2, Declinations -40.0$^{\circ}$ to -26.0$^{\circ}$. Department of Astronomy, University of Michigan, Ann Arbor, MI
\bibitem[Houk \& Smith-Moore(1988)]{houk1988} Houk N., Smith-Moore M., 1988, Michigan Catalogue of Two-dimensional Spectral Types for the HD Stars, Vol. 4, Declinations -26.0$^{\circ}$ to -12.0$^{\circ}$. Department of Astronomy, University of Michigan, Ann Arbor, MI
\bibitem[Howell et al.(2014)]{kepler2} Howell, S. B., Sobeck, C., Haas, M. et al., 2014, PASP, 126, 398

\bibitem[Ida \& Lin(2005)]{Ida2005} Ida, S. \& Lin, D. N. C. 2005, ApJ, 626, 1045
\bibitem[Ida et al.(2013)]{ida2013} Ida, S., Lin, D. N. C. \& Nagasawa, M. 2013, ApJ, 775, 42


\bibitem[Jenkins et al.(2008)]{jenkins2008} Jenkins, J. S., Jones, H. R. A., Pavlenko, Y., et al. 2008, A\&A, 485, 571
\bibitem[Jenkins et al.(2015)]{jenkins2015} Jenkins, J. M., Twicken, J., D., Batalha, N. M., et al. 2015, ApJ, 150, 56
\bibitem[Johnson et al.(2006)]{johnson2006b} Johnson, J.~A., Marcy, G.~W., Fischer, D.~A., et al.\ 2006, \apj, 652, 1724 
\bibitem[Johnson et al.(2007a)]{johnson2007a} Johnson, J.~A., Fischer, D.~A., Marcy, G.~W., et al.\ 2007, \apj, 665, 785 
\bibitem[Johnson et al.(2007b)]{johnson2007b} Johnson, J. A., Butler, R. P., Marcy, G. W., et al. 2007, ApJ, 670, 833
\bibitem[Johnson et al.(2010)]{johnson2010} Johnson, J.~A., Aller, K.~M., Howard, A.~W., \& Crepp, J.~R.\ 2010, \pasp, 122, 905 
\bibitem[Jones et al.(2011)]{jones2011} Jones, M. I., Jenkins, J. S., Rojo, P. \& Melo, C. H. F. 2011, A\&A, 536, 71
\bibitem[Jones et al.(2014)]{jones2014} Jones, M. I., Jenkins, J. S., Bluhm, P., Rojo, P. \& Melo, C. H. F. 2014, A\&A, 566, 113
\bibitem[Jones et al.(2015a)]{jones2015a} Jones, M. I., Jenkins, J. S., Rojo, P.,  Melo, C. H. F. \& Bluhm, P. 2015, A\&A, 573, 3
\bibitem[Jones et al.(2015b)]{jones2015b} Jones, M. I., Jenkins, J. S., Rojo, P., Olivares, F. \& Melo, C. H. F 2015a, A\&A, 580, 14
\bibitem[Jones et al.(2015a)]{jones2015a} \bibitem[Jones et al.(2016)]{jones2016} Jones, M. I., Jenkins, J. S., Brahm, R. et al. 2016, A\&A, 590, 38
\bibitem[Jones et al.(2017)]{jones2017} Jones, M. I., Brahm, R., Wittenmyer, R. A. et al. 2017, A\&A, 602, 58
\bibitem[Jones et al.(2018)]{jones2018} Jones, M. I., Brahm, R., Espinoza, N. et al. 2018, A\&A, 613, 76
\bibitem[Jones et al.(2019)]{jones2019} Jones, M. I., Brahm, R., Espinoza, N. et al. 2019, A\&A, 625, 16

\bibitem[Kaufer et al.(1999)]{kaufer1999} Kaufer, A., Stahl, O., Tubbesing, S. et al. 1999, The Messenger 95, 8
\bibitem[Kennedy \& Kenyon(2008)]{kennedy2008} Kennedy, G. M. \& Kenyon, S. J. 2008, ApJ, 673, 502
\bibitem[Kennedy \& Kenyon(2009)]{kennedy2009} Kennedy, G. M. \& Kenyon, S. J. 2009, ApJ, 695, 1210

\bibitem[Kjeldsen \& Bedding(1995)]{kjeldsen1995} Kjeldsen, H. \& Bedding, T. R. 1995, A\&A, 293, 87
\bibitem[Kokuba \& Ida(1998)]{kokubo1998} Kokubo, E. \& Ida, S. 1998, Icarus, 131, 171
\bibitem[Kunitomo et al.(2011)]{kunitomo2011} Kunitomo, M., Ikoma, M., Sato, B. et al. 2011, \apj, 737, 66 


\bibitem[Laughlin et al.(2004)]{laughlin2004} Laughlin, G., Bodenheimer, P. \& Adams, F. C. 2004, ApJ, 612, L73
\bibitem[Lagrange et al.(2009)]{lagrange2009} Lagrange, A.-M., Desort, M., Galland, F., Udry, S. \& Mayor, M. 2009, A\&A 495, 335

\bibitem[Marcy et al.(2000)]{marcy2000} Marcy, G., Butler, R. P., Fischer, D. A. \& Vogt, S. S., ASPC, 213, 85
\bibitem[Mawet et al.(2014)]{mawet14} Mawet, D., Milli, J., Wahhaj, Z. et al. 2014, ApJ, 792, 97
\bibitem[Mayor \& Queloz(1995)]{mayor1995} Mayor, M. \& Queloz, D: 1995, Nature, 378, 355
\bibitem[M\'elendez et al.(2012)]{Melendez2012} Mel\'endez, J., Bergemann, M., Cohen J.~G. et al. 2012, A\&A, 543, 29
\bibitem[Meschiari et al.(2009)]{MES09} Meschiari, S., Wolf, A. S., Rivera, E. et al. 2009, \pasj, 121, 1016
\bibitem[Moe \& Kratter(2019)]{moe2019} Moe, M. \& Kratter, K. 2019, arXiv: 191201699 
\bibitem[Morton(2015)]{Morton2015} Morton, T. D. 2015, ascl soft, ascl:1503.010
\bibitem[Mulders et al.(2015)]{mulders2015} Mulders, G. D., Pascucci, I. \& Apai, D. 2015, ApJ, 814, 130
\bibitem[Muzerolle et al.(2005)]{muzerolle2005} Muzerolle, J., Luhman, K. L., Brice\~no, C., Hartmann, L., \& Calvet, N. 2005, ApJ, 625, 906


\bibitem[Nelder \& Mead(1965)]{NelderMead65} Nelder, J. A. \& Mead, R. 1965, Computer Journal, 7, 308
\bibitem[Niedzielski et al.(2007)]{niedzielski2007} Niedzielski, A., Konacki, M., Wolszczan, A. et al. 2007, ApJ, 669, 1354

\bibitem[Papaloizou \& Lin(1984)]{papaloizou1984} Papaloizou, J. \& Lin, D. N. C. 1984, ApJ, 285, 818
\bibitem[Petigura et al.(2013)]{petigura2013} Petigura, E. A. Howard, A. W. \& Marcy, G. W. 2013, PNAS; 110, 19273
\bibitem[Pinsonneault et al.(1989)]{yrec} Pinsonneault, M. H., Kawaler, S. D., Sofia, S., \& Demarque, P. 1989, ApJ, 338, 424
\bibitem[Pojmanski(2002)]{asas} Pojmansi, G. 2002, AcA, 52, 397
\bibitem[Pollack et al.(1996)]{pollack1996} Pollack, J. B., Hubickyj, O., Bodenheimer, P. et al. 1996, Icarus, 124, 62

\bibitem[Queloz et al.(2001)]{queloz2001} Queloz, D., Henry, G. W., Sivan, J. P. et al. 2001, A\&A, 379, 279

\bibitem[Reffert et al.(2015)]{reffert2015} Reffert, S., Bergmann, C., Quirrenbach, A., Trifonov, T. \& K\"{u}nstler, A. 2015, A\&A 574, 116
\bibitem[Ricker et al.(2015)]{tess} Ricker, G. R., Winn, J. N., Vanderspek, R., et al. 2015, Journal of Astronomical Telescope, Instruments and Systems, 1, 014003


\bibitem[Sahlmann et al.(2011)]{SAH11} Sahlmann, J., S\'egransan, D., Queloz, D., et al. 2011, A\&A, 525, 95
\bibitem[Sato et al.(2005)]{sato2005} Sato, B., Kambe, E., Takeda, Y. et al. 2005, PASJ, 57, 97
\bibitem[Sato et al.(2008)]{sato2008} Sato, B., Toyota, E., Omiya, M. et al. 2008, PASJ, 60, 1317
\bibitem[Scargle(1982)]{scargle82} Scargle, J. D. 1982, ApJ, 263, 835
\bibitem[Setiawan et al.(2003)]{setiawan2003} Setiawan, J., Pasquini, L., da Silva, L., von der L\"{u}he, O.\& Hatzes, A. 2003, A\&A, 397, 1151
\bibitem[Sneden et al.(2012)]{moog} Sneden, C., Bean, J., Ivans, I. et al. 2012, ascl soft, ascl:1202.009
\bibitem[Soto \& Jenkins(2018)]{species} Soto, M. \& Jenkins, J. S. 2018, A\&A, 615, 76
\bibitem[Stassun \& Torres(2018)]{stassun2018} Stassun, K. G. \& Torres, G. 2018 ApJ, 862, 61
\bibitem[Stello et al.(2009)]{stello2009} Stello, D., Chaplin, W. J., Basu, S.,  Elsworth, Y. \& Bedding, T. R. 2009, MNRAS, 400, 80
\bibitem[Stello et al.(2017)]{stello2017} Stello, D.,  Huber, D.,  Grundahl, F. et al. 2017, MNRAS, 472, 4110

\bibitem[Tokovinin et al.(2013)]{TOK13} Tokovinin, A., Fischer, D. A., Bonati, M. et al. 2013, PASP, 125, 1336
\bibitem[Thorngren et al.(2016)]{thorngren2016} Thorngren, D.~P., Fortney, J.~J.; Murray-Clay, R.~A. \& Lopez, E.~D. 2016, ApJ, 831, 64
\bibitem[Trifonov et al.(2019)]{Trifonov2019_es} Trifonov, Trifon, 2019, (ascl:1906.004)

\bibitem[van Leeuwen(2007)]{hipparcos2007} van Leeuwen, F. 2007, A\&A, 474, 653
\bibitem[Villaver \& Livio(2009)]{villaver2009} Villaver, E. \& Livio, M. 2009, \apj, 705, 81

\bibitem[Winn et al.(2009)]{winn2009} Winn, J. N., Johnson, J. A., Albrecht, S. et al. 2009, ApJ, L99, 103
\bibitem[Wittenmyer et al.(2011)]{wit2011} Wittenmyer, R. A.; Endl, M., Wang, L. et al. 2011, ApJ, 743, 184
\bibitem[Wittenmyer et al.(2016a)]{wit2016a} Wittenmyer, R. A., Johnson, J. A., Butler, R. P. et al. 2016, ApJ, 818, 35
\bibitem[Wittenmyer et al.(2016b)]{wit2016b} Wittenmyer, R. A., Liu, F., Wang, L. et al. 2016, AJ, 152, 19
\bibitem[Wittenmyer et al.(2017a)]{wit2017a} Wittenmyer, R. A., Jones, M. I., Zhao, J. et al. 2017a, AJ, 153, 51
\bibitem[Wittenmyer et al.(2017b)]{wit2017b} Wittenmyer, R.~A., Jones, M.~I., Horner, J., et al.\ 2017b, \aj, 154, 274 
\bibitem[Wittenmyer et al.(2020a)]{wittenmyer2020a} Wittenmyer, R.~A., Wang, S.,  Horner, J. et al. 2020a, MNRAS, 492, 377
\bibitem[Wittenmyer et al.(2020b)]{wittenmyer2020b} Wittenmyer, R.~A., Butler, R. P., Horner, J. et al. 2020b, MNRAS, 491, 5248

\bibitem[Zechmeister \& K\"{u}rster(2009)]{Zechmeister2009} Zechmeister \& K\"{u}rster 2009, A\&A, 496, 577


\end{thebibliography}
\end{document}